# An illustrative experiment on electromagnetic oscillations

J. Escalante, J. L. Vázquez, D. Montes and L. E. López
Natural Science Department
Inter American University of Puerto Rico
Bayamón Campus, Bayamón PR00957
(jescalante@bc.inter.edu)

When measuring electrical voltage or electrical current in electrical circuits with analog meters, digital instruments or an oscilloscope, the user ought to be sensitive to readings of brief electrical signals or pulses in response to minute changes in the electrical components or compatibility with the internal resistance of meter devices. There are situations when laboratory measurements are embarrassing initially although prove later to be instructional.

Measuring of time constant of capacitors connected in series with a DC power source and a resistor (known as RC circuit) is a standard exercise in any university physics course in natural science [1-4]. So that, it is intensely discuss the exponential behavior of charging and discharging of capacitors both in the conference and in the laboratory setting. Later, it is discussed the presence of a coil with inductance, a resistor and a capacitor all connected in series with a DC power source (known as RLC circuit) but rarely shown in the laboratory exercises [4]. It is the purpose of this manuscript to place an illustrative demonstration on the measurement of damped electromagnetic oscillations for a RLC circuit that it is easy to set in any physics laboratory equipped with PASCO technologies and USB Electrical PASPort sensors together with standard electrical components. The results of recording the electrical voltage with DATA Studio software [5] have a very good agreement with performed simulations from MULTISIM software [6] and/or standard calculations from theory. Our students and instructors enjoy of the experiment for their simplicity set up in addition to the instructive oscillations.

Our experimental set up is depicted in figure (1). There you can see the circuit boards with resistances, coil inductors, capacitors, wires, a power amplifier (PA) with a voltage sensor (VS) plugged in, a Science Workshop Interface (SWI) 750 box, and computer monitor.

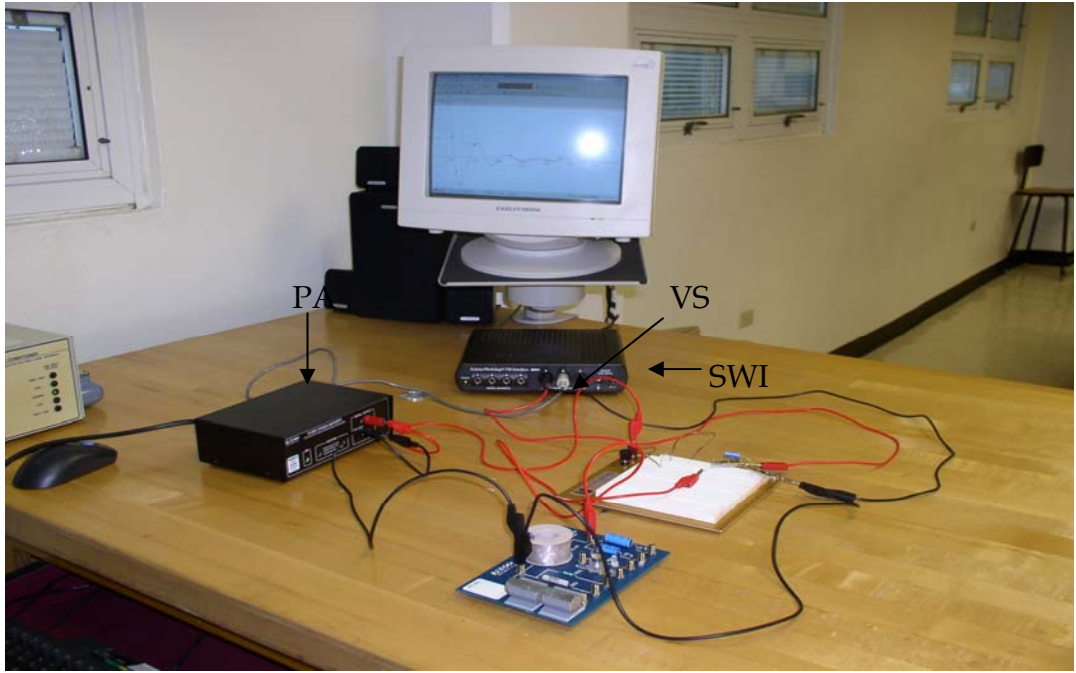

*Fig. 1. This is a picture of the actual RLC circuit set up for the viewing of the electromagnetic damped oscillations.*

A schematic diagram of the experimental set up is in figure (2). There, it is shown two circuits, one that we call *the secondary circuit* and shall be used to completely charge the capacitor and the other circuit to discharge the capacitor and this circuit we call it *the primary circuit*. It is in this primary circuit in which it is monitored the electromagnetic oscillations by measuring the voltage signals from the capacitor. Alternatively, this can be made by measuring the counter electromotive force through the inductor.

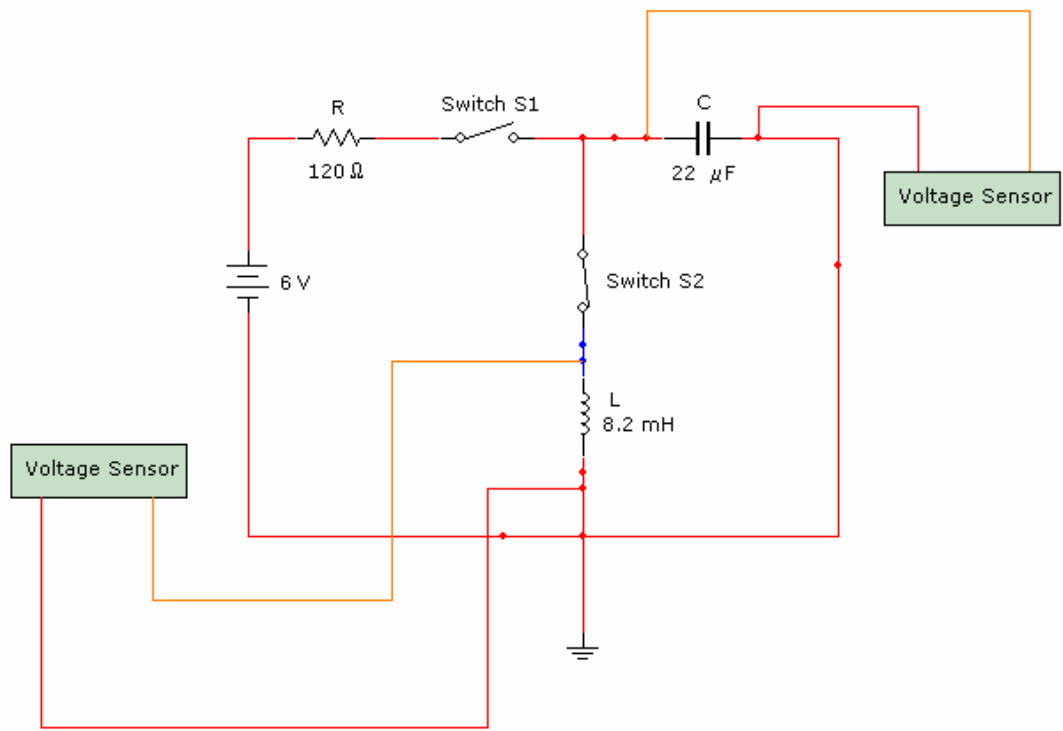

*Fig. 2. This schematic shows the required connections with primary and secondary circuits to observe the electromagnetic oscillations in this experiment.*

The secondary circuit is made of a 120 ohms resistor, a 22 microfarads capacitor and 6 volts DC power supply together with a switch S1 that is connected in between the resistance and the capacitor. While the capacitor is being charge the switch S1 is kept close and switch S2 that is part of the primary circuit is kept opened. The complete charging of the capacitor it takes about two seconds. The primary circuit is made of a 22 microfarads capacitor, an 8.2 millihenries inductor together with a switch S2 that is connected in between the capacitor and the inductor. While the capacitor is being discharged the switch S2 is kept close and switch S1 that is part of the secondary circuit is kept opened. The complete discharging of the capacitor takes about half a second. It is during this time interval that the electromagnetic oscillations occur.

Soon after the capacitor has been charged, switch S1 opened and switch S2 closed, from energy considerations it follows that the rate of change of the

electromagnetic energy in the primary circuit is given by the following expression:

$$\frac{d}{dt}(U_B + U_E) = -i^2 R, \qquad (1)$$

where $U_B$ is the intensity of the magnetic energy for the coil with inductance (L) and $U_E$ is the intensity of the electric energy for the capacitor (C). In time, the electromagnetic energy will be dissipated by the internal resistance (R) of the electric circuit. The above expression can be shown to reduce to the following familiar differential equation:

$$\frac{d^2q}{dt^2} + \frac{R}{L}\frac{dq}{dt} + \frac{1}{LC}q = 0. \qquad (2)$$

A solution for any of the above expressions which shows explicitly damped voltage (V) oscillations requires:

$$V = V_0\, e^{-\left(\frac{R}{2L}t\right)} \cos(\omega t + \delta) \qquad (3)$$

Where, $V_0$ is the voltage of the fully charged capacitor, $\delta$ a phase angle and $\omega$ is the frequency for the damped oscillations

$$\omega = \sqrt{\frac{1}{LC} - \frac{R^2}{4L^2}} \qquad (4)$$

We run the experiment as follows: The electric energy was supplied to our electrical circuit by a power amplifier CI-6552A from PASCO plugged into an analog channel A. The voltage was set to 6 volts DC. It was placed a resistance of 120 Ohms in series with a Switch S1 and a capacitor of 22 microfarads to completely charge it. Then, wired the capacitor in series with a switch S2 and a coil inductance of 8.2 millihenries, to have a closed loop without the power amplifier source. The switches S1 and S2 made easy of switching between primary and secondary circuits. We placed a voltage sensor model CI-6503 from PASCO at low sensitivity and sample rate to 10 kHz, connected to a science interface 750 box in channel B, to monitor voltage across the capacitor. From the measurements recorded with Data Studio software we have values of 6.0 volts at complete charged capacitor and in switching secondary to primary circuit to measure the voltage oscillations through capacitor we have values like 5.850 volts at

3.8796 seconds, -2.949 volts at 3.8809 seconds, 1.484 volts at 3.8822 seconds, -0.772 volts at 3.8835 seconds, 0.410 volts at 3.8847 seconds, and so on for higher picks in voltage damped oscillations. It is important to set the millisecond scale in time to view these values and shape of curve, otherwise we can only see a step down crossing the zero reference and continuing in zero voltage reading. Our results are shown in figure (3).

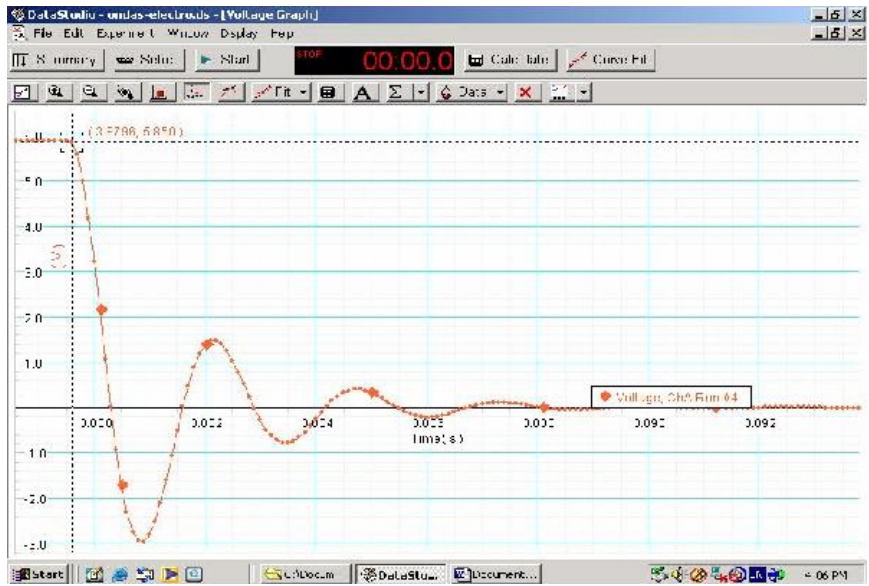

*Fig. 3. This illustration from DATA STUDIO shows electrical oscillations for the LC circuit with values for capacitance 22 microfarads and inductance of 8.3 millihenries. The viewing of oscillations is set in the millisecond scale.*

Using nominal values for the capacitor (22microfarads), the inductor (8.2 millihenries) and time and voltage corresponding to the maxima and minima of the oscillations, see figure 3, together with the expected theoretical behavior for the damped voltage oscillations we found a value for the internal resistance of the circuit to be $(8.1 \pm 0.3)$ohms.

This computed value for the internal resistance together with nominal values of capacitor; inductor and a DC power source at 5.85 volts were used to run a MULTISIM simulation of this experiment. The schematic of the simulated circuit is shown in figure (4) and the result of the simulation is depicted in figure (5)

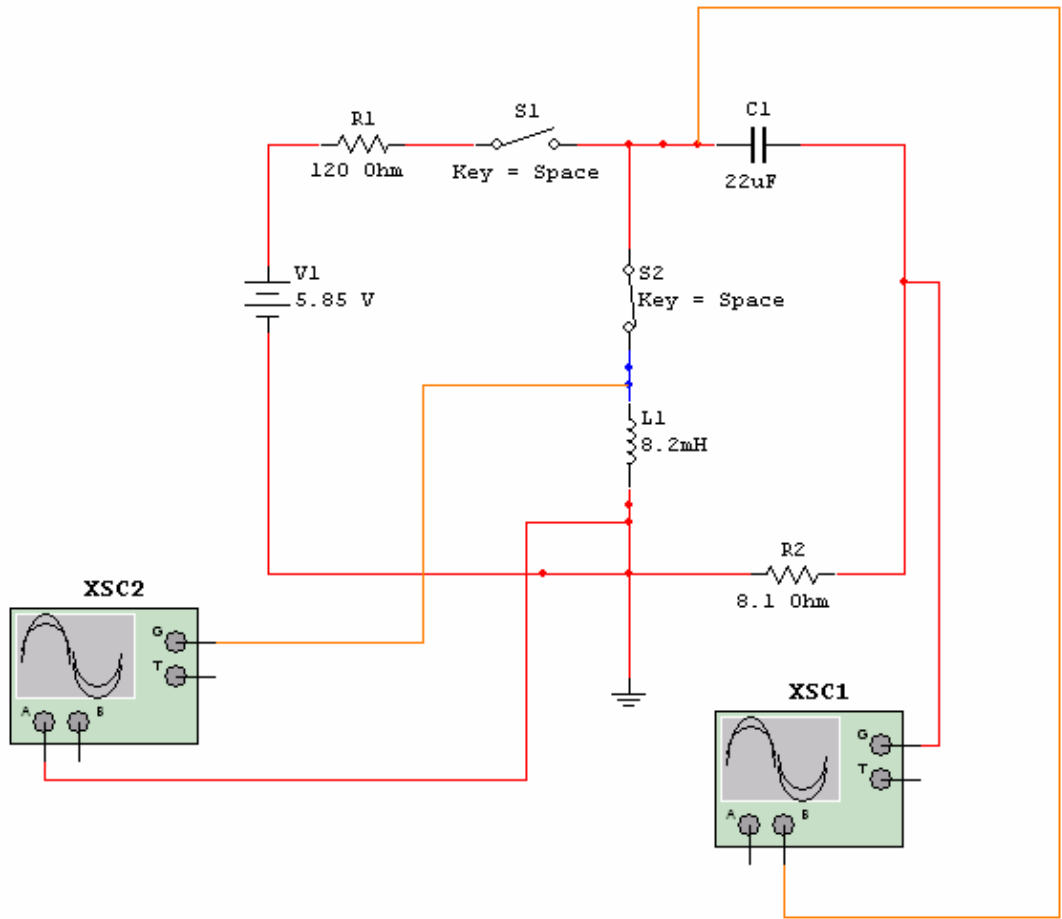

*Fig. 4. This is a MULTISIM simulation for the schematic of the RLC circuit connected in series with a power source. XSC1 and XSC2 are oscilloscopes to monitor voltage in capacitor and inductor, respectively.*

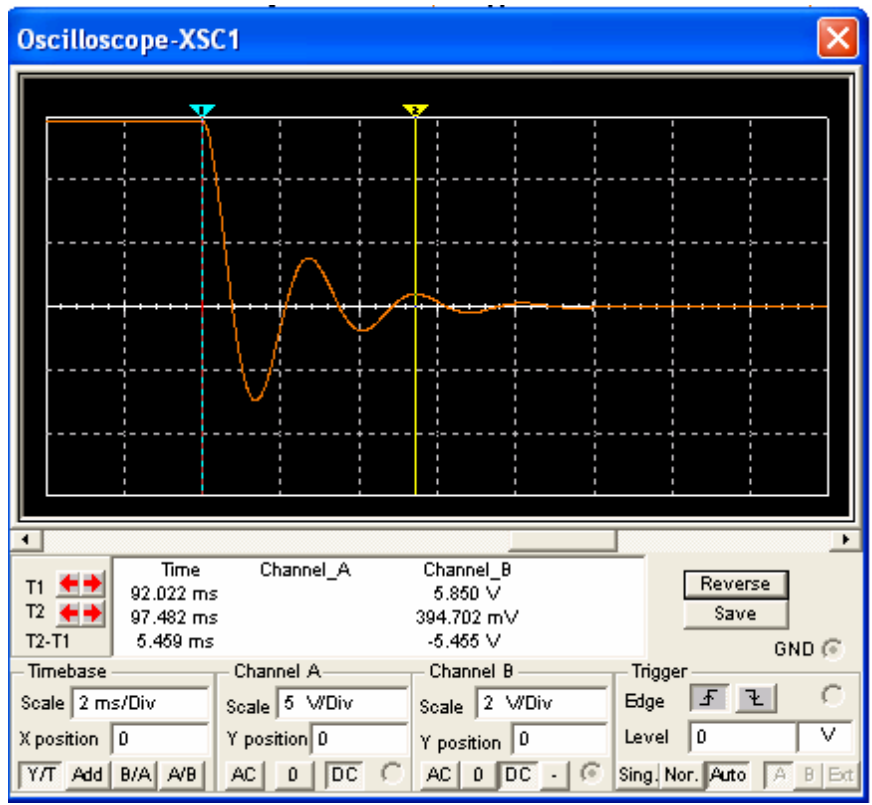

*Fig. 5. This is a MULTISIM simulation for electromagnetic voltage oscillations as read directly from XSC1 oscilloscope. The parameters of the oscillations are easy to read in the front panel controls.*

A comparison of the experimental theoretical and MULTISIM simulation results with the maxima and minima values for voltage are presented in table 1.

| Time (milliseconds) | | | Voltage (volts) | | |
|---|---|---|---|---|---|
| experimental | theoretical | simulation | experimental | theoretical | Simulation |
| 0.0 | 0.0 | 0.0 | 5.85 | 5.85 | 5.85 |
| 1.3 | 1.340 | 1.378 | -2.949 | -3.013 | -2.981 |
| 2.6 | 2.681 | 2.755 | 1.484 | 1.547 | 1.518 |
| 3.9 | 4.021 | 4.031 | -0.772 | -0.791 | -0.764 |
| 5.1 | 5.362 | 5.459 | 0.410 | 0.404 | 0.395 |

*Table 1. Comparison of time and voltage values corresponding to maxima and minima of the oscillations between experimental, theoretical and simulated results.*

The results we have from this simulation are in good agreement with values from theoretical calculations and also with real experimental measurements.

As can be seen for table 1, the oscillating period as measured experimentally is 2.55 milliseconds is only 5% percentage difference with theoretical value. Similarly the percentage difference between experimental and simulated values is within 7%. Taking into account that we used the nominal values for capacitor and inductor together with the average value for the internal resistance of the circuit we conclude that above experiment in very stimulating to students and instructors because of the very good agreement between the experimental results and expected theoretical behavior of the oscillating circuit. For the above reasons it is highly recommended for the introductory physics curses to place an experiment on the measurement of electromagnetic oscillations and set up for the physics lectures of an interactive lecture demonstration [7].

Biographies:

J. Escalante is professor of physics in the mathematics and science department at Inter American University of Puerto Rico since 1994. He graduated in nuclear physics from University of South Carolina in 1989. jescalante@bc.inter.edu

J.L. Vazquez is adjunct professor in the mathematics and science department since 2005. He graduated in Mexico and received his Ph.D. from QMW College, London England. jlbellox@netscape.net

D. Montes Kercadó is the physics laboratory technician in the natural science department at Inter American University of Puerto Rico, Bayamón Campus. He obtained his B.S. in theoretical physic from University of Puerto Rico, Mayagüez Campus, Mayagüez, P.R. dmontes@bc.inter.edu

L. E. López Colón will graduate in 2005 as an electrical engineering from Inter American University of Puerto Rico. He is majoring in communications. Luiso_el_loco@hotmail.com